\documentclass[aps,prc,twocolumn,superscriptaddress]{revtex4}

\usepackage{amssymb}
\usepackage{amsmath}
\usepackage{lscape}
\usepackage{booktabs}

\usepackage{graphicx}
\usepackage{epstopdf}
\usepackage{graphics}
\usepackage{graphicx}
\usepackage{epsfig}
\usepackage{epstopdf}

\graphicspath{{figures/}}

\newcommand{\dd}{\ensuremath{{\rm d}}}
\newcommand{\pT}{\ensuremath{p_{\rm T}}}
\newcommand{\Ncoll}{\ensuremath{N_{\rm coll}}}
\newcommand{\Npart}{\ensuremath{N_{\rm part}}}
\newcommand{\TAA}{\ensuremath{T_{\rm AA}}}

\newcommand{\sNN}{\ensuremath{\sqrt{s_{\rm NN}}}}
\newcommand{\s}{\ensuremath{\sqrt{s}}}

\newcommand{\dndy}{\ensuremath{\dd N/\dd y}}

\newcommand{\Zn}{\ensuremath{{\rm Z}^{0}}}
\newcommand{\Wpm}{\ensuremath{{\rm W}^{\pm}}}
\newcommand{\Wp}{\ensuremath{{\rm W}^{+}}}
\newcommand{\Wm}{\ensuremath{{\rm W}^{-}}}
\newcommand{\WZ}{\ensuremath{\Wpm / \Zn}}

\newcommand{\mupm}{\ensuremath{{\rm \mu}^{\pm}}}
\newcommand{\mup}{\ensuremath{{\rm \mu}^{+}}}
\newcommand{\mum}{\ensuremath{{\rm \mu}^{-}}}

\newcommand{\PYTHIA}{\textsc{Pythia}}

\newcommand{\cent}[2] {$#1$--$#2\%$}
\newcommand{\avg}[1]{\left\langle #1 \right\rangle}

\usepackage[dvipsnames]{xcolor}

\begin{document}

\title{Dynamical simulation on production of $\Wpm$ and $\Zn$ bosons in \\
p--p, p--Pb (Pb--p), and Pb--Pb collisions at $\sNN=5.02$~TeV with PACIAE}
\author{Dai-Mei Zhou$^1$ \footnote{zhoudm@mail.ccnu.edu.cn},
Yu-Liang Yan$^{1,2}$ \footnote{yanyl@ciae.ac.cn}, Liang Zheng$^{3}$,
Ming-Rui Zhao$^2$, Xiao-Mei Li$^2$, Xiao-Ming Zhang$^1$, Gang Chen $^{3}$,
Xu Cai$^1$, and Ben-Hao Sa$^{1,2}$ \footnote{sabh@ciae.ac.cn}}
\affiliation{$^1$ Key Laboratory of Quark and Lepton Physics (MOE) and
Institute of Particle Physics, Central China Normal University, Wuhan 430079,
China.\\
  $^2$ China Institute of Atomic Energy, P. O. Box 275 (10), Beijing, 102413 China. \\
  $^3$ School of Mathematics and Physics, China University of Geosciences (Wuhan), Wuhan 430074, China.}

\begin{abstract}
In this paper, production of $\Zn$ and $\Wpm$ vector bosons in p--p, p--Pb (Pb--p), and Pb--Pb collisions at
$\sNN=5.02$~TeV is dynamically simulated with a parton and hadron cascade model PACIAE. ALICE data of $\Zn$
production is found to be reproduced fairly well. A prediction for $\Wpm$ production is given in the same
collision systems and at the same energy. An interesting isospin-effect is observed in the sign-change of
$\mupm$ charge asymmetry in pp, pn, np, and nn collisions and in minimum bias p--Pb, Pb--p and Pb--Pb
collisions at $\sNN=5.02$~TeV, respectively.
\end{abstract}

\maketitle

\section {Introduction}
$\Wpm$ and $\Zn$ vector bosons are heavy particles with masses of $m_{\Wpm}=80.39$~GeV/$c^{2}$ and $m_{\Zn}=91.19$~GeV/$c^{2}$~\cite{pdg}.
They are mainly produced in the large momentum transferred hard partonic scattering processes at the early stage of the (ultra-)relativistic
nuclear-nuclear collisions. Their main production processes are
\begin{equation*}
u\overline{d}\rightarrow\Wp, \hspace{0.5cm} d\overline{u}\rightarrow\Wm
\end{equation*}
and
\begin{equation*}
u\overline{u}\rightarrow\Zn, \hspace{0.5cm} d\overline{d}\rightarrow\Zn
\end{equation*}
in leading order approximation \cite{martin}.
Therefore, the different abundance ratios of valence quarks $u$ and $d$ in p--p, p--Pb (Pb--p), and Pb--Pb collisions may result in
difference on the ratio of $\Wp$ and $\Wm$ production yields among those systems. This is the so called isospin effect.

In comparing with evolution time of the heavy-ion collision system, $10$ to $100$~fm/$c$ for instance, decay time of $\WZ$, which can be
estimated with the full decay width $\Gamma$ \cite{pdg},
\begin{eqnarray*}
t = \frac{\hbar}{\Gamma},\hspace{0.4cm}
t_{\Wpm} = 0.0922~{\rm fm/}c,\hspace{0.4cm}
t_{\Zn} = 0.0791~{\rm fm/}c,
\end{eqnarray*}
is very short.
The $\WZ$ leptonic decays
\begin{equation*}
\Wp\rightarrow l^{+}\nu_{l}, \hspace{0.5cm} \Zn\rightarrow l^{+}l^{-},~(l{\rm :}~e,\mu,\tau)
\end{equation*}
are nearly instantaneous. As the produced leptons weakly interact with the partonic and hadronic matters, $\Wpm$ and $\Zn$, similar as the prompt direct photons, are powerful probes for investigating the properties of the initial stage of the evolving system and the partonic structure of the colliding nuclei.

Aforementioned properties are in the microscopic sector. In the macroscopic part, for heavy-ion collisions, the problems to be addressed are the geometric properties of the two colliding nuclei overlapped region with a given impact parameter $b$. In this sector, the key parameters are the nuclear thickness function $\avg{\TAA}$ (angle bracket denotes the average over events), the number of participant nucleons $\avg{\Npart}$, and the number of binary collisions $\avg{\Ncoll}$.
They are calculated using the Glauber model \cite{shor,abel,abel1,misk}, in which the relation of
\begin{equation}
\avg{\Ncoll}=\sigma_{\rm NN}^{\rm inel}\times\avg{\TAA}
\label{tanc}
\end{equation}
is important.

The CMS and ATLAS Collaborations have first measured $\Wpm$ and $\Zn$ production in Pb--Pb collisions at $\sNN=2.76$~TeV~\cite{cms1,atlas1,cms2,atlas2}.
Recently, the ALICE and ATLAS Collaborations published the measurements of $\Zn$ production at forward rapidities~\cite{alice1} and $\WZ$ production at mid-rapidity~\cite{atlas3,atlas4}, in Pb--Pb collisions at $\sNN=5.02$~TeV, respectively. A similar measurement of $\Wpm$ production in Pb--Pb collisions with ALICE is on the way. The $\Wpm$ and $\Zn$ production cross sections are also measured in p--Pb collisions at $\sNN=5.02$~and/or $8.16$~TeV with ALICE and CMS~\cite{alice2,cms3}. All those measurements are declared to be well reproduced by the leading-order (LO) and next-to-leading-order (NLO) perturbative Quantum Chromo Dynamics (pQCD) calculations~\cite{ct14,eps09,cteq15,epps} using the CT14 Parton Distribution Function (PDF) set \cite{ct14} with and without the parameterized nuclear modified PDF (nPDF) like EPPS16~\cite{epps}. As the experimental data analysis relies on templates calculated with LO pQCD, comparing experimental data to the LO or NLO pQCD predictions is incomprehensive. The study of $\WZ$ production in heavy-ion collisions with dynamical simulation may provide more differential understandings into the microscopic transport properties of the partonic system.

\section {Model}
A parton and hadron cascade model PACIAE~\cite{sa1} is employed in this paper to dynamically simulate $\Zn$ production in p--p and Pb--Pb collisions at
$\sNN=5.02$~TeV. The results are compared with that measured by ALICE~\cite{alice1}. Production of $\Wpm$ is predicted in the p--p, p--Pb (Pb--p),
and Pb--Pb collisions at $\sNN=5.02$~TeV as well.

\begin{widetext}
\begin{center}
\begin{figure}[htbp]
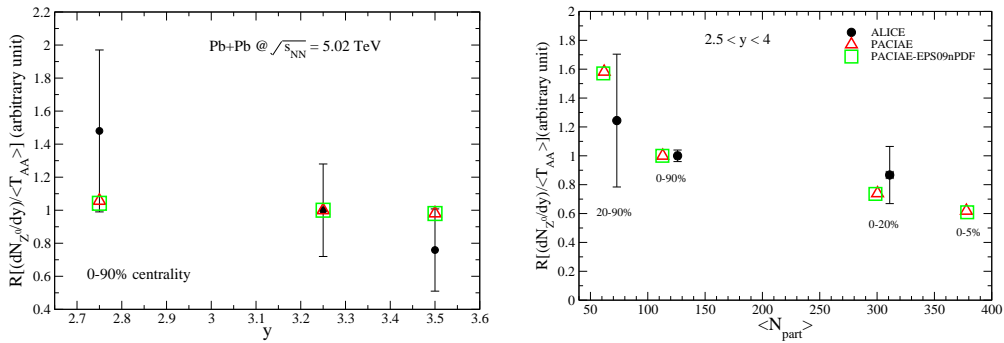

\centering
\hspace{-0.50cm}
\includegraphics[width=0.36\textwidth]{RdndyZ0-y.eps}\hspace{0.5cm}
\includegraphics[width=0.35\textwidth]{RdndyZ0-npart.eps}
\caption{Left panel is the rescaled $\Zn$ rapidity density ($\dndy$~/~$\avg{\TAA}$) as a function of rapidity ($y$). In this panel the results are shown in three $y$ intervals [$2.5,~3]$, $[2.5,~4]$ and $[3,~4]$. The points are presented at the center of each of the intervals. Right panel shows the centrality dependence of R($\dndy/\avg{\TAA}$).}
\label{znpart}
\end{figure}
\end{center}
\end{widetext}

The PACIAE model is based on \PYTHIA\ event generator (version $6.4.28$)~\cite{soj1}. For pp collisions, with respect to \PYTHIA, the partonic and hadronic rescatterings are introduced in PACIAE, before string formation and after the hadronization, respectively. The final hadronic states are developed from the initial partonic hard scatterings followed by the parton and hadron rescattering stages. Thus, the PACIAE model provides a multi-stage transport description on the evolution of the collision system.

For heavy-ion collisions, the initial positions of nucleons in the colliding nucleus are described by the Woods-Saxon distribution and the number of
participant (spectator) nucleons calculated by the Glauber model~\cite{shor,abel,abel1,misk}. Together with the initial momentum setup of $p_{x} = p_{y} = 0$ and $p_{z} = p_{\rm beam}$ for each nucleon, a list containing the initial state of all nucleons in a given nucleus--nucleus colliding system is constructed.
A collision happened between two nucleons if their relative transverse distance is less than or equal to the minimum approaching distance:
$D\leq\sqrt{\sigma_{\rm NN}^{\rm tot}/\pi}$. The collision time is calculated with the assumption of straight-line trajectories. All such nucleon pairs
compose a nucleon-nucleon (NN) collision (time) list. A NN collision with least collision time is selected from the list and executed by \PYTHIA\
(PYEVNW subroutine) with the hadronization temporarily turned-off and the strings as well as diquarks broken-up. The nucleon list and NN collision list
are then updated. A new NN collision with least collision time is selected from the updated NN collision list and executed with repeating the
aforementioned step until the NN collision list is empty.

With those procedures, the initial partonic state for a nucleus-nucleus collision is constructed. Then it proceeds into a partonic rescattering stage
where the LO-pQCD parton-parton cross section~\cite{ranft,field} is employed. After partonic rescatterings, the string is recovered and then hadronized with
the Lund string fragmentation regime resulting in an intermediate hadronic state. Finally, the system proceeds into the hadronic rescattering stage and results in the final hadronic state of the collision system.

The $\WZ$ production yield is very low, e.g., $\dd N(\Zn) / \dd y\sim 10^{-9}$ at mid-rapidity in the most $20\%$ central Pb--Pb collisions at $\sNN=
5.02$~TeV. In our simulations, the relevant production channels are activated in a user controlled approach by setting MSEL$=0$ together with the following
subprocesses switched on:
\begin{eqnarray*}
f_{i}\overline{f}_{j} & \rightarrow & \Wp/\Wm   \\
f_{i}\overline{f}_{j} & \rightarrow & g\Wp/\Wm  \\
f_{i}\overline{f}_{j} & \rightarrow & \gamma \Wp/\Wm \\
f_{i}g                & \rightarrow & f_{k}\Wp/\Wm   \\
f_{i}\overline{f}_{j} & \rightarrow & \Zn\Wp/\Wm     \\
f_{i}\overline{f}_{i} & \rightarrow & \Wp\Wm
\end{eqnarray*}
for $\Wpm$ production, and
\begin{eqnarray*}
f_{i}\overline{f}_{i} & \rightarrow & \gamma^{*}/\Zn    \\
f_{i}\overline{f}_{i} & \rightarrow & g(\gamma^{*}/\Zn) \\
f_{i}\overline{f}_{i} & \rightarrow & \gamma(\gamma^{*}/\Zn) \\
f_{i}g                & \rightarrow & f_i(\gamma^{*}/\Zn)    \\
f_{i}\overline{f}_{i} & \rightarrow & (\gamma^{*}/\Zn)(\gamma^{*}/\Zn) \\
f_{i}\overline{f}_{j} & \rightarrow & \Zn\Wp/\Wm
\end{eqnarray*}
for $\Zn$ production. In aforementioned equations $f$ refers to fermions(quarks) and its subscript stands for flavor code.

As $\Wpm$ and $\Zn$ bosons are nearly transparency in both the Quark Gluon Matter and Hadron Matter, they are not considered in the partonic rescattering
and hardonic rescattering in the PACIAE simulations. Thus the results of $\Wpm$ and $\Zn$ productions from PACIAE simulations are nearly the same as
the ones from PYTHIA simulations for p--p collisions.

\begin{widetext}
\begin{center}
\begin{figure}[htbp]
\centering
\hspace{-0.50cm}
\includegraphics[width=0.25\textwidth,angle=270]{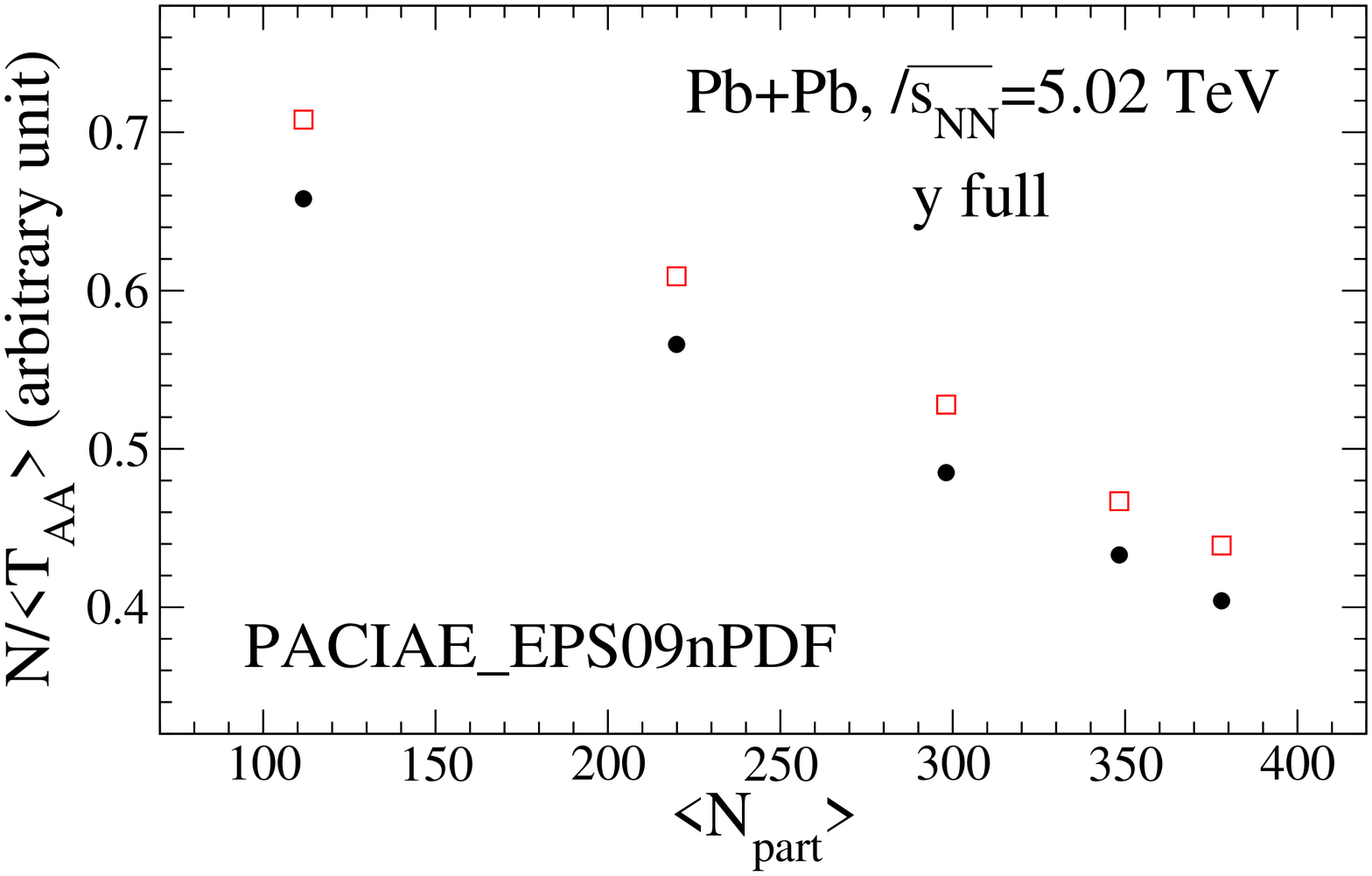}\hspace{0.02cm}
\includegraphics[width=0.25\textwidth,angle=270]{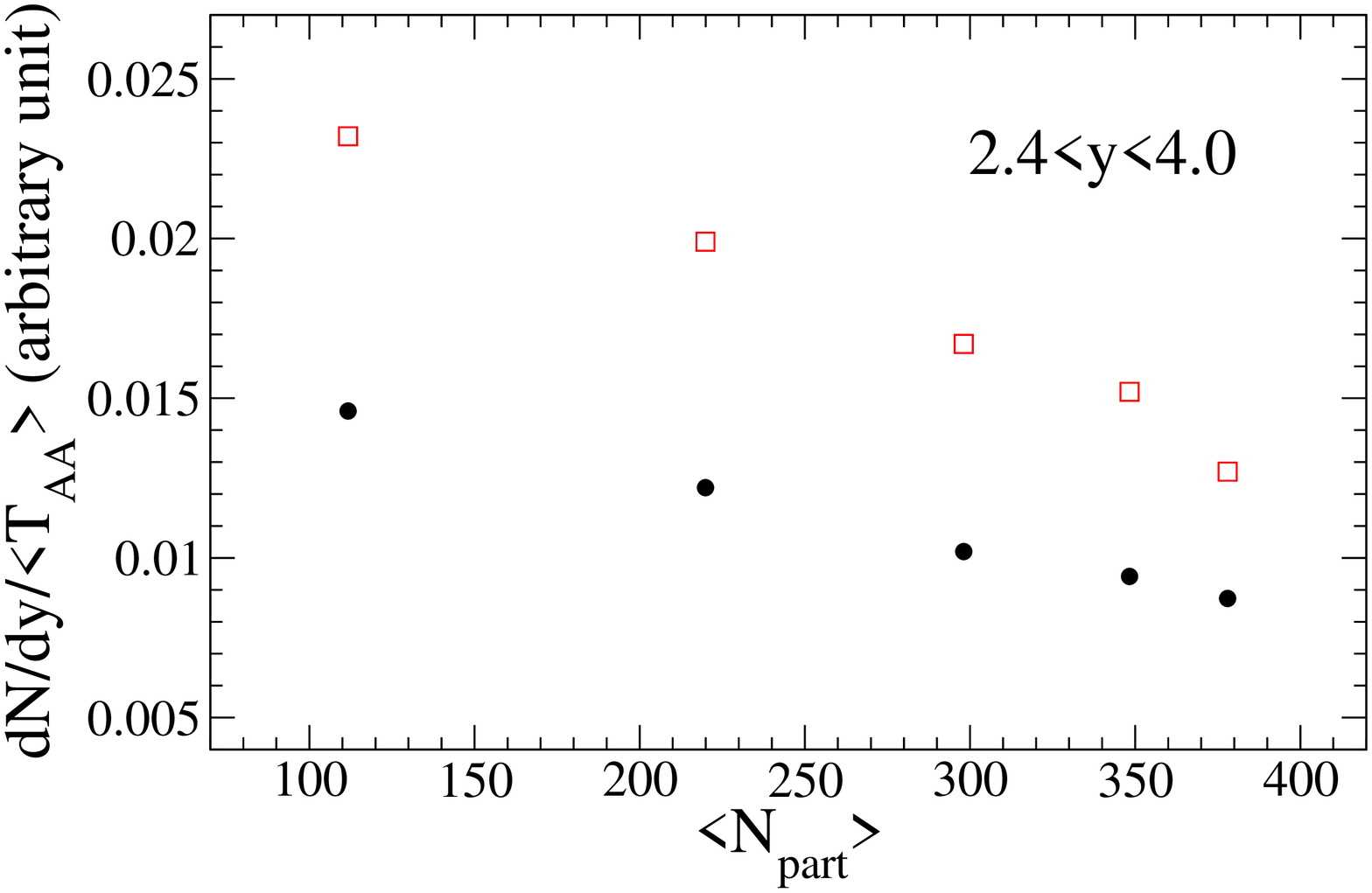}\hspace{0.02cm}
\includegraphics[width=0.25\textwidth,angle=270]{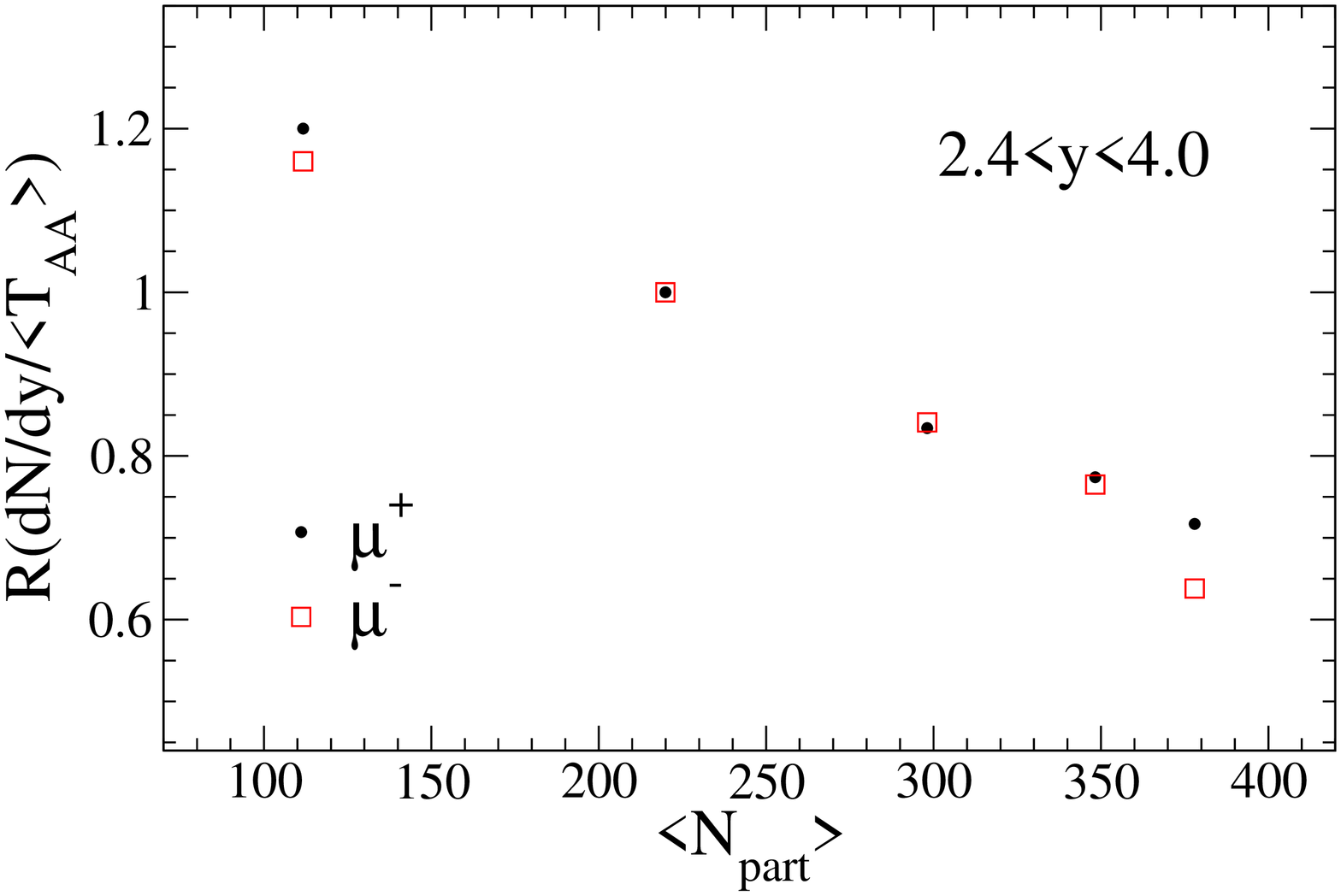}
\caption{The centrality dependence of $N/\avg{\TAA}$ (left panel), $\dndy/\avg{\TAA}$ (middle panel), and
$R(\dndy/\avg{\TAA})$  (right panel) for the $\mup$ and $\mum$ decaying from $\Wp$ and $\Wm$, respectively, in the Pb--Pb collisions at
$\sNN=5.02$~TeV simulated with PACIAE-EPS09nPDF.}
\label{munpar}
\end{figure}
\end{center}
\end{widetext}

The Monte Carlo simulation of $\WZ$ production described above is a triggered
$\WZ$ production approach (a bias sampling technique). A normalization factor
is needed to account for the trigger bias effect (the bias correction). To
make a fair comparison to the experimental data, we use the rescaled
distribution defined as follows
\begin{equation}\label{eq:res}
R(X) = X / X_{\rm ref},
\end{equation}
where $X$ denotes a given observed distribution, such as the rapidity density $\dndy$~/~$\avg{\TAA}$.
Here $X_{\rm ref}$ is a chosen reference point in the distribution. The comparison between data and simulations will be presented on the rescaled
distribution $R(X)$.

\begin{widetext}
\begin{center}
\begin{figure}[htbp]
\centering
\hspace{-0.50cm}
\includegraphics[width=0.3\textwidth,angle=270]{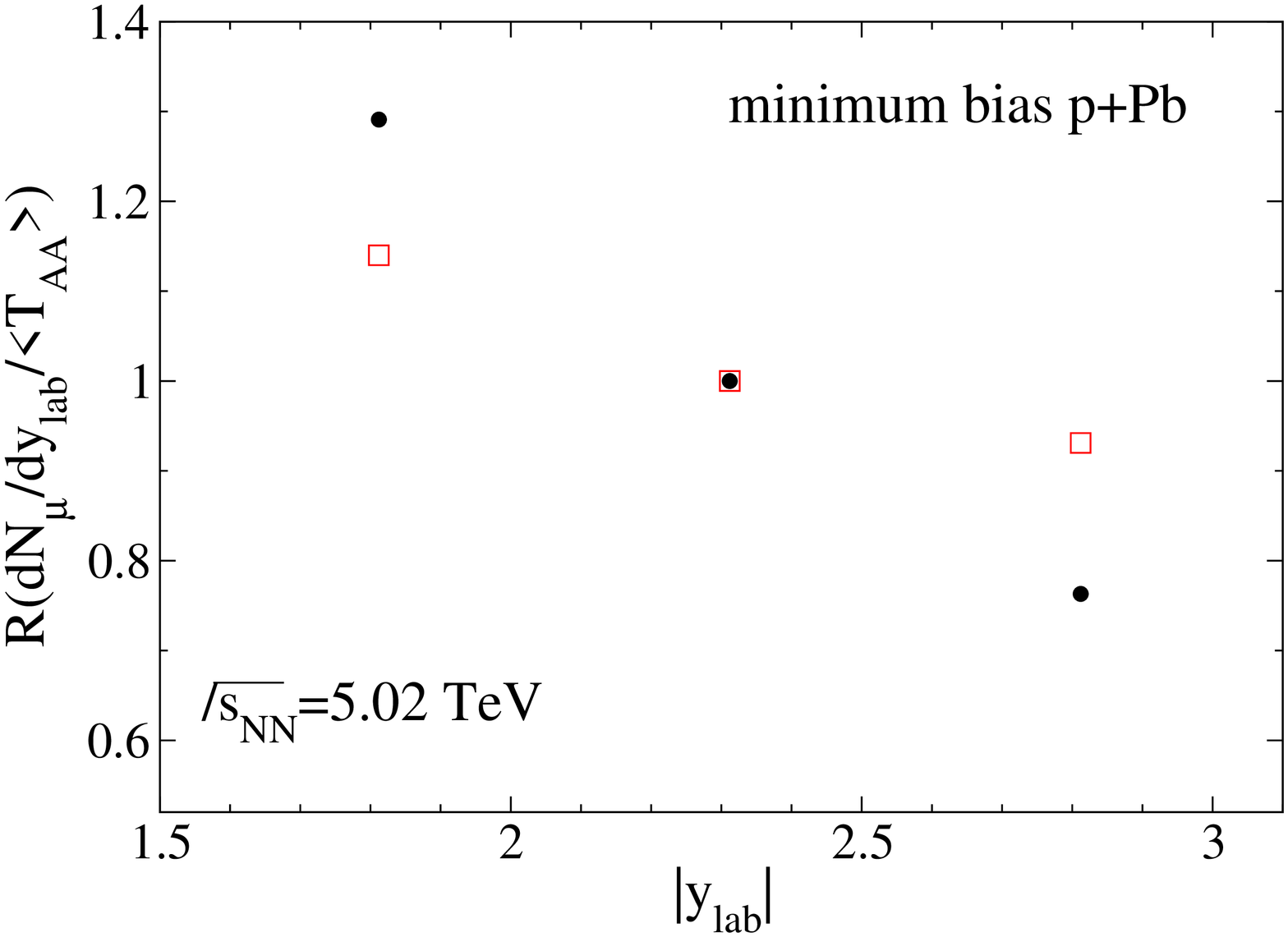}\hspace{0.05cm}
\includegraphics[width=0.3\textwidth,angle=270]{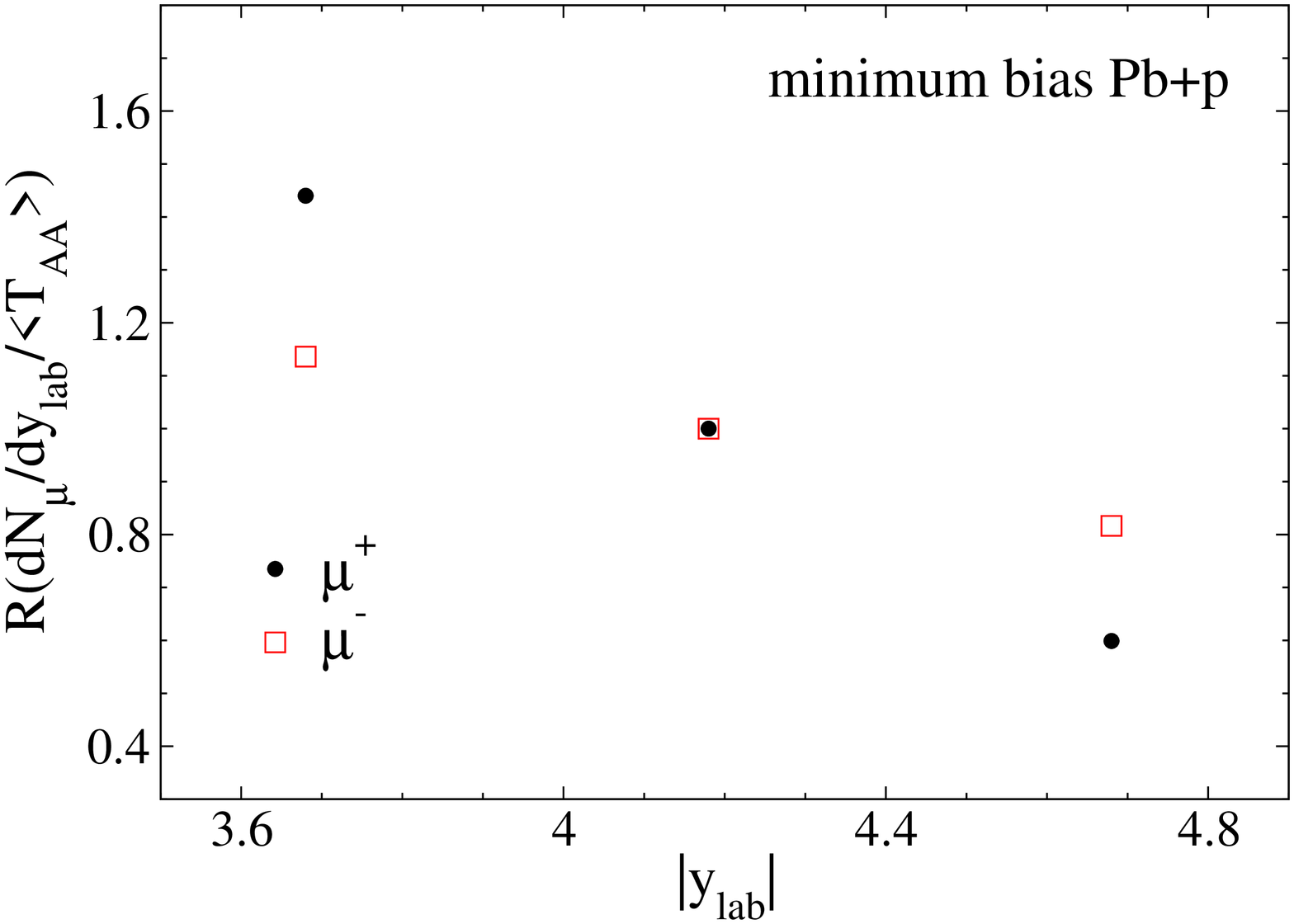}
\caption{The rescaled $\mu$ rapidity density ($dN/dy/<T_{AA}>$) as a function of $|y_{lab}|$, for $\mup$ and $\mum$ decayed from
$\Wp$ and $\Wm$, in minimum bias p--Pb (left panel) and Pb--p (right panel) collisions at $\sNN=5.02$ ~TeV respectively.}
\label{ppb_dnmuy}
\end{figure}
\end{center}
\end{widetext}

\section {Results and Discussions}
The comparison of the rescaled $\Zn$ rapidity-differential density $R(\dd N_{\Zn}/\dd y~/~\avg{\TAA})$ between PACIAE simulations and the ALICE
measurements is shown in the left panel of Fig.~\ref{znpart} for \cent{0}{90} centrality class in Pb--Pb collisions at $\sNN=5.02$~TeV. The points on the
plot, from the left to right, represent the results in rapidity intervals of $2.5 < y < 3$, $2.5 < y < 4$ and $3 < y < 4$. In both data and simulations,
the value in $2.5 < y < 4$ is chosen as the reference point. In this figure, the black full circles are the ALICE measurements ~\cite{alice1}, the red open triangles are PACIAE results with free proton PDF, and the green open squares are PACIAE results with EPS09 nPDF~\cite{eskola}. This panel shows that the ALICE measurements~\cite{alice1} are well reproduced by PACIAE dynamical simulations within uncertainties.

The right panel of Fig.~\ref{znpart} shows the centrality dependent of rescaled $R(\dd N_{\Zn}/\dd y/\avg{\TAA})$ in the Pb--Pb collisions at $\sNN=5.02$~TeV. The red open triangles are PACIAE results with free proton PDF, and the green open squares are those with EPS09nPDF, while the ALICE data are indicated by the black full circles. Again, the ALICE data are well reproduced within error bars.

\begin{widetext}
\begin{center}
\begin{figure}[htbp]
\hspace{-0.50cm}
\includegraphics[width=0.25\textwidth,angle=270]{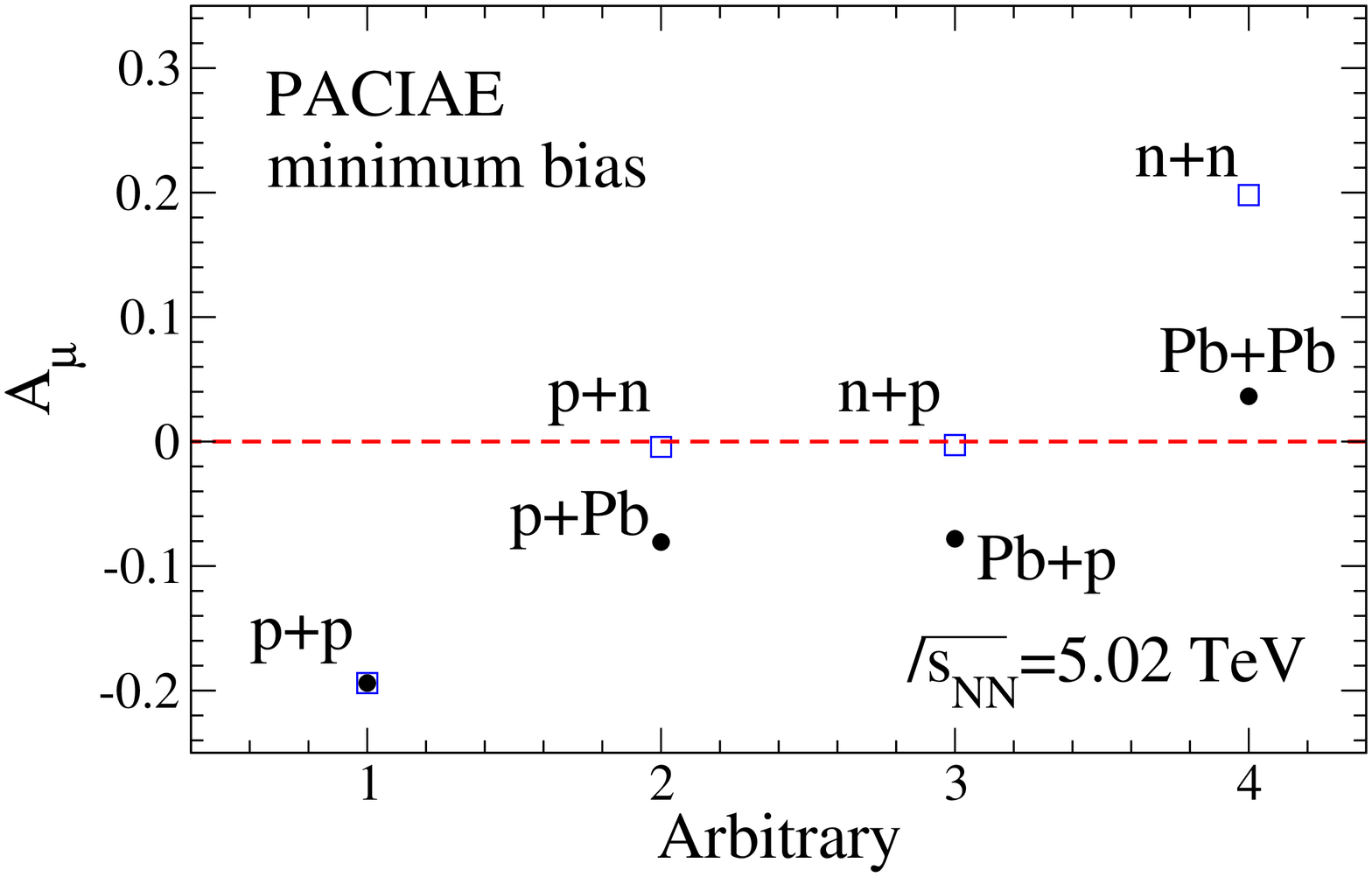}\hspace{0.02cm}
\includegraphics[width=0.25\textwidth,angle=270]{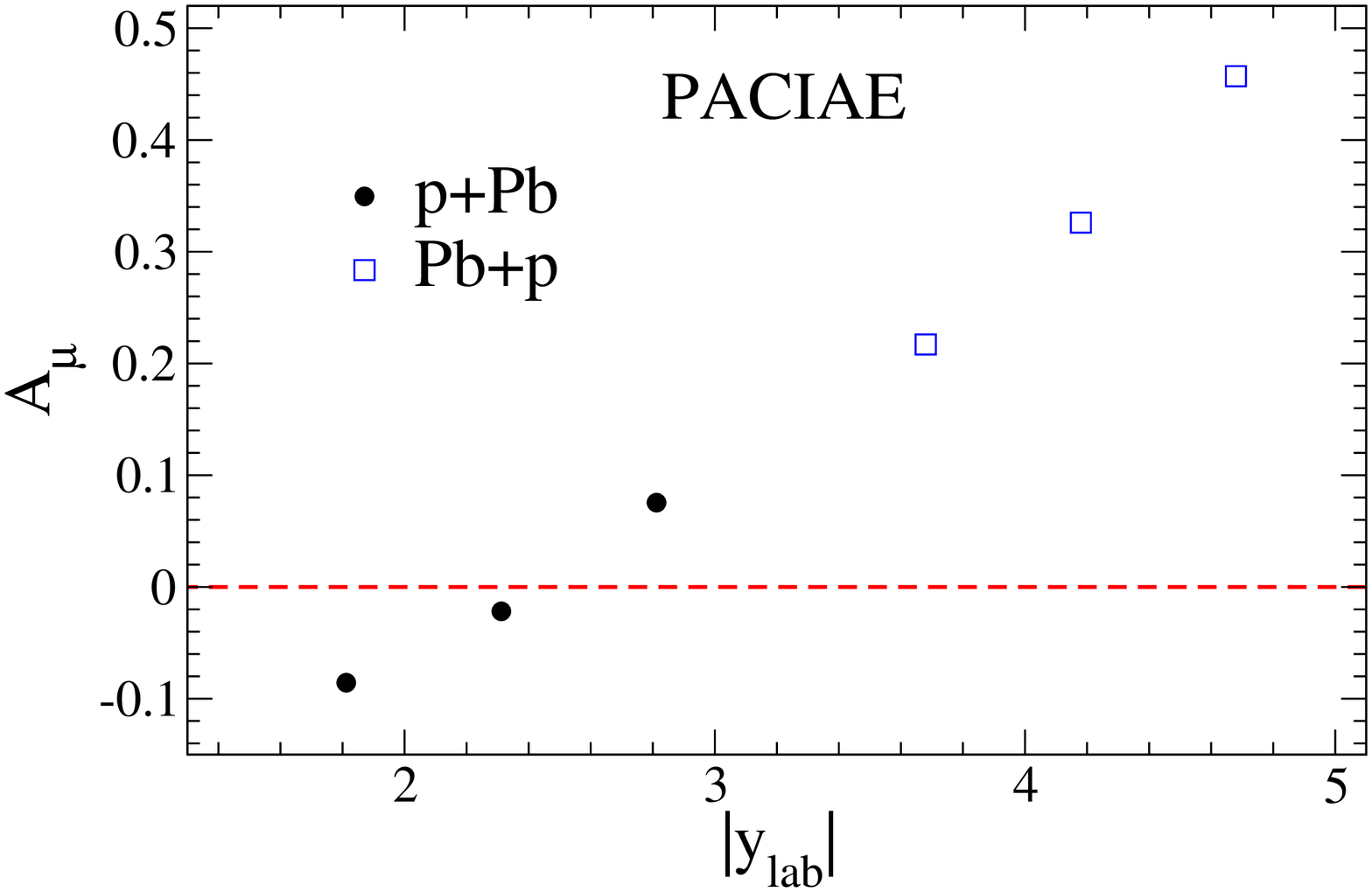}\hspace{0.02cm}
\includegraphics[width=0.25\textwidth,angle=270]{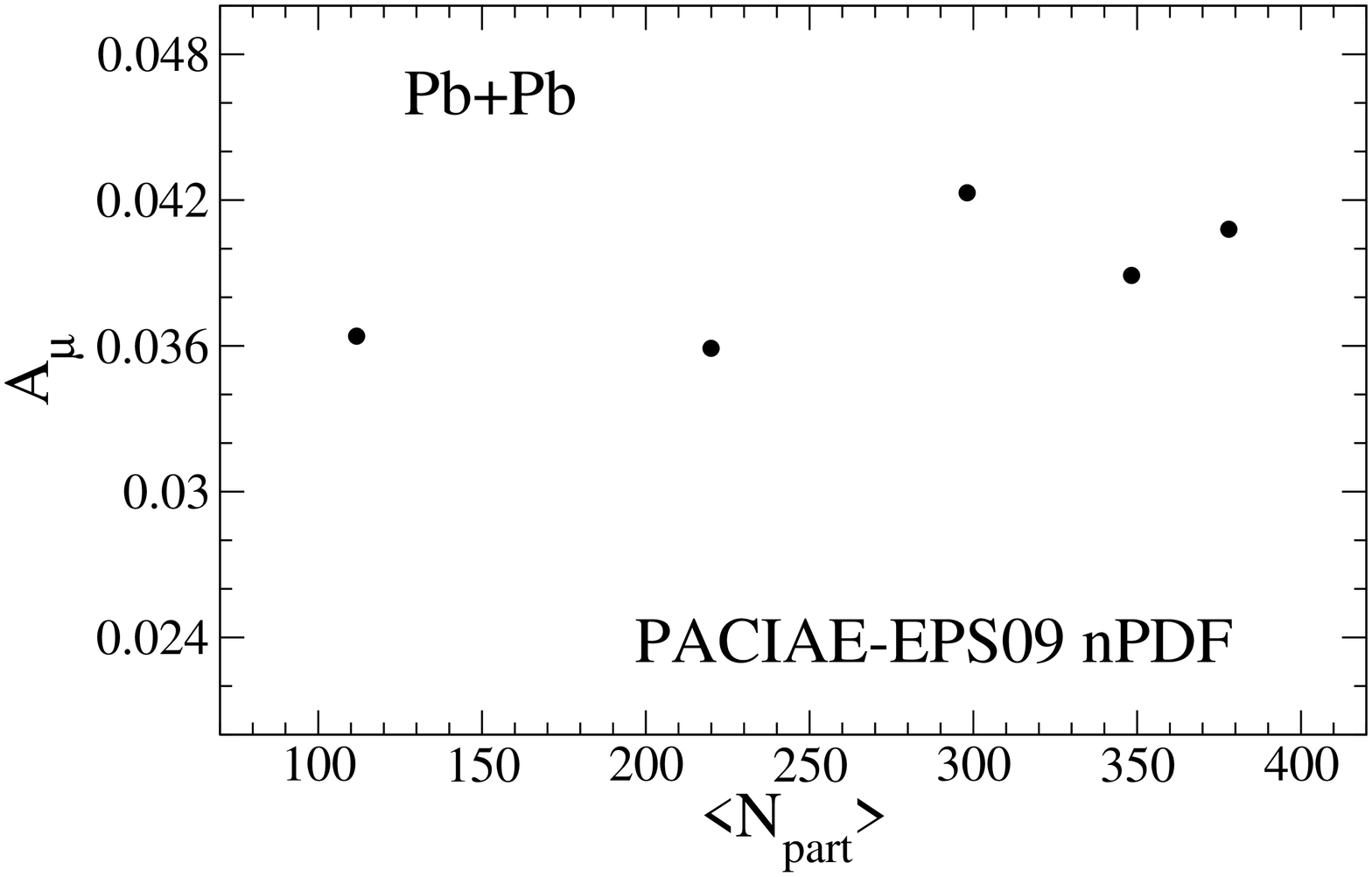}
\caption{The $\mu$ charge asymmetry in minimum bias p--p, p--n, n--p, and n--n collisions at center of energy equal to 5.02 TeV is given in left panel with blue open squares, from left to right respectively. The black full circles in this panel are the same as blue ones but for minimum bias p--p, p--Pb, Pb--p
and Pb--Pb collisions at $\sqrt{s_{NN}}$=5.02 TeV, respectively. middle panel is the $\mu$ differential charge asymmetry in minimum bias p--Pb (black full
circles) and Pb--p (blue open squares) collisions at $\sqrt{s_{NN}}$ =5.02 TeV. The $\mu$ charge asymmetry as a function of $<\rm N_{part}>$ in Pb--Pb
collisions at $\sqrt{s_{NN}}$=5.02 TeV is given in right panel.}
\label{amu}
\end{figure}
\end{center}
\end{widetext}

Similar model calculations for $\mupm$ production in Pb--Pb collisions at $\sNN=5.02$~TeV are shown in Fig.~\ref{munpar}. In this figure and in
following studies for Pb--Pb collisions, the calculations are made in event centralities \cent{0}{5}, \cent{0}{10}, \cent{0}{20}, \cent{0}{40}, and
\cent{0}{90} within the corresponding impact parameter intervals of $[0,~3.5]$, $[0,~4.94]$, $[0,~6.98]$, $[0,~9.88]$ and $[0,~14.96]$~fm to match the event
geometries shown in ALICE data~\cite{abel1}. The $\avg{\Npart}$ from optical Glauber calculations are, correspondingly, $378.6$, $348.3$, $298.1$, $219.9$
and $111.7$.

This figure shows the centrality dependent $N/\avg{\TAA}$ (left panel), the corresponding $\dndy/\avg{\TAA}$ (in $2.4 < y < 4$, middle panel) and the rescaled
distribution of $R(\dndy/\avg{\TAA})$ (in $2.4 < y < 4$, right panel) obtained from PACIAE simulations with nPDF for $\Wpm$ decay $\mupm$ in Pb--Pb collisions at
$\sNN=5.02$~TeV. The trends of the distributions shown in the left and middle panels are similar to what are shown in Fig.~$13$ in~\cite{atlas3}. However the $\mupm$ is obtained in the full $\pT$ phase space in PACIAE simulations while ATLAS measures that in
$\pT>25$~GeV/$c$, where the event reliability is very low.

Meanwhile, we present the model calculation for the rescaled distributions, which is defined in eq.~(\ref{eq:res}), for $\mupm$ from $\Wpm$ decays in
p--Pb and Pb--p collisions at $\sNN=5.02$~TeV in Fig.~\ref{ppb_dnmuy}. The results are presented in rapidity
intervals of $[2.03,~2.53]$, $[2.53,~3.03]$, and $[3.03,~3.53]$ in p--Pb collisions, and of $[2.96,~3.46]$, $[3.46,~3.96]$, and $[3.96,~4.46]$ in Pb--p
collisions. A rapidity shift $\Delta y=0.465$ has been used to account for the asymmetric beam energy configurations. In these figures, different rapidity
dependence for $\mup$ and $\mum$ are observed between the proton-going and Pb-going directions.

The asymmetry between $\Wp$ and $\Wm$ production yields, stemming from the isospin effect, can be studied with the asymmetry of their decay products
$\mup$ and $\mum$ as follows
\begin{equation}
A_{\rm \mu}=\frac{Y_{\mum}-Y_{\mup}}{Y_{\mum}+Y_{\mup}}.
\end{equation}
The $\mupm$ charge asymmetries in minimum bias pp, pn, np, and nn collisions at $\s=5.02$~TeV are presented in the left panel of Fig.~\ref{amu} as the blue open squares. The black full circles are the results form minimum bias p--Pb, Pb--p, and Pb--Pb collisions at $\sNN=5.02$~TeV. In the middle panel of
Fig.~\ref{amu}, we present the differential charge asymmetry as a function of $|y_{lab}|$ of $\mupm$ in minimum bias p--Pb (black full circles) and Pb--p
(blue open squares) collisions at $\sNN=5.02$~TeV. The rapidity intervals presented here are the same as that in Fig.~\ref{ppb_dnmuy}.
The right panel of Fig.~\ref{amu} shows the $\mupm$ asymmetry varying with $\avg{\Npart}$ in Pb--Pb collisions at $\sNN=5.02$~TeV.

The hierarchy of the charge asymmetry observed in pp, pn, np and nn collisions shown in Fig.~\ref{amu} can be understood by the variation of relative
abundance of the valence $u$ and $d$ quarks in those colliding hadron objects. The similar trend observed with nuclear collision beams thus arises due to the
increasing neutron abundance from pp to Pb--Pb collisions. The Pb-beam is more neutron-like than the proton-beam. Therefore, the sign flipping of $A_{\mu}$ from p--Pb to Pb--p collisions shown in the middle panel of Fig.~\ref{amu} is a natural outcome of the different valence kinematic dominance between proton-side and Pb-side detector acceptance regions.

\begin{widetext}
\begin{center}
\begin{figure}[htbp]
\centering
\hspace{-0.50cm}
\includegraphics[width=0.75\textwidth,angle=270]{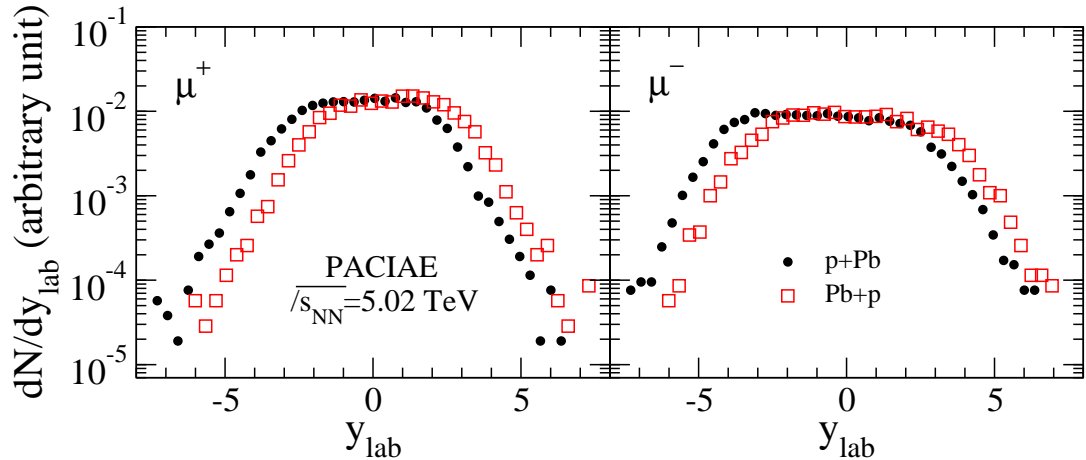}
\caption{The left and right panels are rapidity distributions of $\mup$ and $\mum$ decayed from $\Wp$ and $\Wm$, respectively. They are dynamically simulated
in minimum bias p--Pb and Pb--p collisions at $\sNN=5.02$~TeV by PACIAE model.}
\label{y_ppbpbp}
\end{figure}
\end{center}
\end{widetext}

The PACIAE simulations on rapidity distributions of $\mup$ (left panel) and $\mum$ (right panel) in the minimum bias p--Pb and Pb--p collisions at
$\sNN=5.02$~TeV are given in Fig.~(\ref{y_ppbpbp}). The relations of $y_{lab}=y_{cms}-0.465$ and $y_{lab}=y_{cms}+0.465$ are used in p--Pb and Pb--p collisions, respectively.

Figure.~(\ref{ymu}) shows the PACIAE (with EPS09 nPDF) simulations on rapidity distributions of $\mup$ (left panel) and $\mum$ (right panel) for \cent{0}{10}
(black full circles) and \cent{0}{90} (red open squares) in Pb--Pb collisions at $\sNN=5.02$~TeV. The results are compared with that in pp (blue open
circles) collisions at the same energy. It shows that the shape of the $y$-differential distribution in \cent{0}{90} Pb--Pb collisions is more
similar to the distribution in pp than to that in the most $10\%$ Pb--Pb. The rapidity plateau of $\mum$ is observed to be wider than $\mup$ in all the
collision systems studied.

\begin{widetext}
\begin{center}
\begin{figure}[htbp]
\centering
\hspace{-0.50cm}
\includegraphics[width=0.75\textwidth,angle=270]{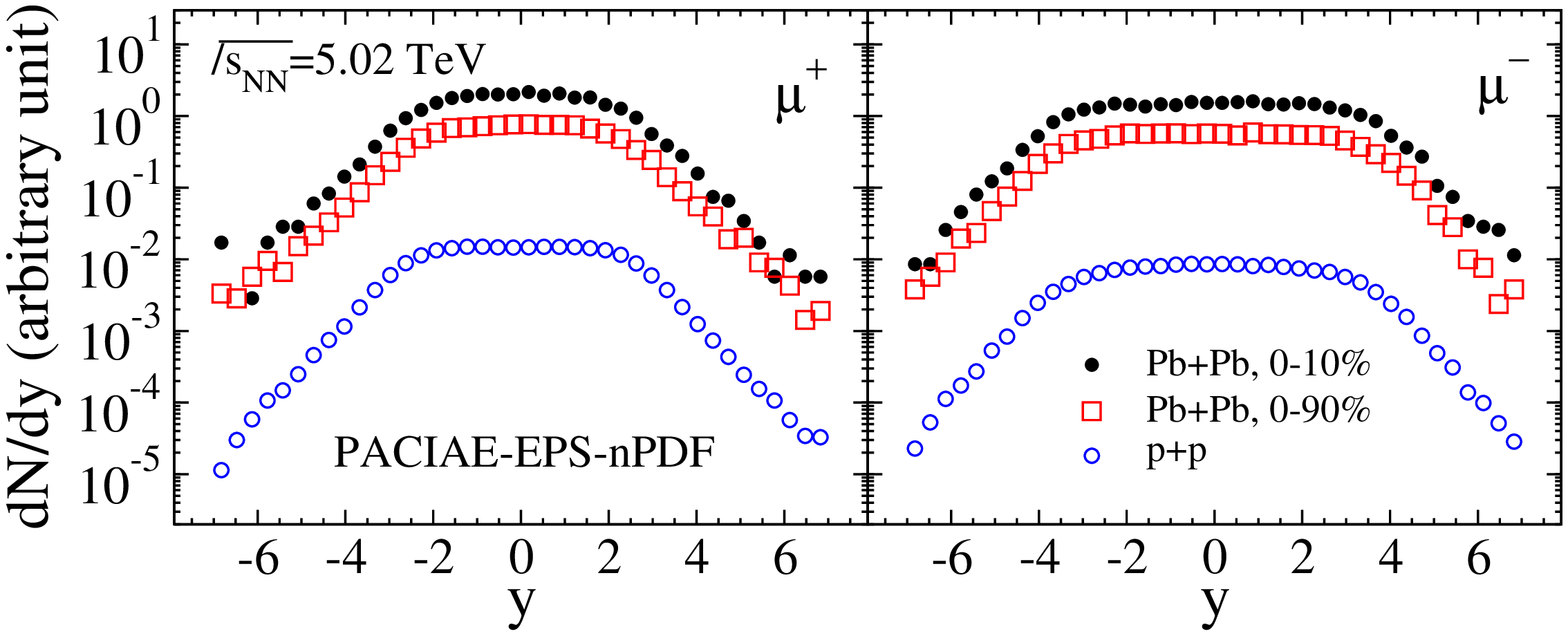}
\caption{The rapidity distribution of $\mup$ (left panel) and $\mum$ (right
panel) decayed from $\Wp$ and $\Wm$, respectively. They are dynamically
simulated in \cent{0}{10} and \cent{0}{90} central classes Pb--Pb as well as
pp collisions at $\sNN=5.02$~TeV by PACIAE model with EPS09nPDF.}
\label{ymu}
\end{figure}
\end{center}
\end{widetext}

\section {Summary and Acknowledgment}
The parton and hadron cascade model of PACIAE is employed simulating the dynamical production of $\WZ$ bosons in pp, p--Pb (Pb--p) and Pb--Pb collisions at $\sNN=5.02$~TeV for first time in this paper. The rescaled $\dndy/\avg{\TAA}$ for $\Zn$ bosons measured by ALICE in Pb--Pb collisions at $\sNN=5.02$~TeV~\cite{alice1} are fairly reproduced. Simulations on $\mupm$ production are given for all collision systems. A sign-change of $\mupm$ charge asymmetry are observed in pp, pn, np, and nn collisions and in minimum  bias p--Pb, Pb--p and Pb--Pb collisions at $\sNN=5.02$~TeV, respectively. These interesting isospin-effect observations are worthwhile to be investigated further. Meanwhile, carrying out studies of $\WZ$ boson production in dynamical simulations with partonic transport effects may shed a light on the understanding of medium induced higher order effects in the future works.

This work was supported by the National Natural Science Foundation of China (11775094, 11805079, 11905188, 11775313),
the Continuous Basic Scientific Research Project (No.WDJC-2019-16) in CIAE, National Key Research and Development
Project (2018YFE0104800) and by the 111 project of the foreign expert bureau of China.


\begin{thebibliography}{}
\bibitem{pdg}
Particle data group, Review of Particle Physics, Chinese Phys. C 38 (2014) 27.
\bibitem{martin}
A. D. Martin, R. G. Roberts, W. J. Stirling and R. S. Thorne, Eur. Phys. J. C 14 (2000) 133, arXiv: hep-ph/9907231 [hep-ph].
\bibitem{shor}
A. Shor and R. Longacre, Phys. Lett. B 218, 100 (1989).
\bibitem{abel}
B. I. Abelev, et al., STAR Collaboration, Phys. Rev. C 79, 034909 (2009).
\bibitem{abel1}
B. I. Abelev, et al., ALICE  Collaboration, Phys. Rev. C 88, 044909 (2013).
\bibitem{misk}
D. Miskowiec,http://www.linux.gsi.de/-misko/overlap/.
\bibitem{cms1}
CMS Collab., Phys. Lett. B 715 (2012) 66, arXiv: 1205.6334 [hep-ex].
\bibitem{atlas1}
ATLAS Collab., Eur. Phys. J. C 75 (2015) 23, arXiv: 1408.4674 [hep-ex].
\bibitem{cms2}
CMS Collab., Phys. Rev. Lett. 106 (2011) 212301, arXiv: 1102.5435 [hep-ex].
\bibitem{atlas2}
ATLAS Collab., Phys. Rev. Lett. 110 (2013) 022301, arXiv: 1210.6486 [hep-ex].
\bibitem{alice1}
ALICE Collab., Phys. Lett. B 780 (2018)372, arXiv: 1711.10753v2 [hep-ex].
\bibitem{atlas3}
ATLAS Collab., Eur. Phys. J. C 79 (2019) 935, arXiv: 1907.10414v1 [hep-ex].
\bibitem{atlas4}
ATLAS Collab., arXiv: 1910.13396v1 [hep-ex].
\bibitem{alice2}
ALICE Collab., JHEP, 02. 077 (2017), arXiv: 1611.03002v2 [hep-ex];
ALICE Collab., arXiv: 2005.11126v1 [hep-ex].
\bibitem{cms3}
CMS Collab., Phys. Lett. B 800 (2020) 135048.
\bibitem{ct14}
S. Dulat, T.-J. Hou, J. Gao, M. Guzzi, J. Huston, P. Nadolsky, J. Pumplin, C. Schmidt, D. Stump, and C. P. Yuan, Phys. Rev. D 93 (2016) 033006, arXiv:1506.07443v2 [hep-ph].
\bibitem{eps09}
K. J. Eskola, H. Paukkunen, and C. A. Salgado, JHEP, 04. 065 (2009),
arXiv: 0902.4154 [hep-ph].
\bibitem{cteq15}
A. Kusina, F. Lyonnet, D. B. Clak, E. Godat, T. Jezo, K. Kovarik, F. I. Olness, I Schienbein, and J. Y. Yu, arXiv: 1610.02925 [nucl-th].
\bibitem{epps}
K. J. Eskola, P. Paakkinen, H. Paukkunen, and C. A. Salgado, Eur. Phys. J. C 77 (2017) 163, arXiv: 1612.05741 [hep-ph].
\bibitem{sa1}
Ben-Hao Sa, Dai-Mei Zhou, Yu-Liang Yan, Xiao-Mei Li, Shene-Qin Feng, Bao-Guo Dong, and Xu Cai., Comput. Phys. Commun. 183, 333 (2012); ibid, 224, 412 (2018).
\bibitem{soj1}
T. S\"ojstrand, S. Mrenna, and P. Skands, JHEP, 05, 026 (2006).
\bibitem{ranft}
B. L. Combridge, J. Kripfgang, and J. Ranft, Phys. Lett. B 70, 234 (1977).
\bibitem{field}
R. D. Field, Application of perturbative QCD, Addison-Wesley Publishing Company,
Inc., 1989.
\bibitem{eskola}
L. Helenius, K. J. Eskola, H. Honkanen, and A. Salgado, JHEP, 07. 073 (2012).
\end{thebibliography}
\end{document}